\documentclass[journal,twoside,twocolumn]{IEEEtran}
\usepackage{amsmath,amssymb}
\usepackage{graphicx}
\usepackage{xcolor}
\usepackage{cite}
\usepackage{balance}
\usepackage{makecell}
\usepackage{tabularx}
\usepackage{enumitem}
\usepackage{subcaption}
\usepackage{soul,comment}
\usepackage{hyperref}
\usepackage{moresize}

\usepackage{tikz}
\usetikzlibrary{positioning,shadows,shapes,fit,arrows,backgrounds}
\usepackage{pgfplots}
\pgfplotsset{compat=1.14} %

\newcommand{\bR}{\boldsymbol{R}}

\newcommand{\bs}{\boldsymbol{s}}

\newcommand{\bZ}{\boldsymbol{Z}}
\newcommand{\by}{\boldsymbol{y}}
\newcommand{\hi}{\hat{\imath}}

\newcommand{\Rc}{R_\mathrm{c}}
\DeclareMathOperator*{\argmin}{arg\,min} %

\newcommand{\new}[1]{{#1}}

\usepackage{bbm}

\title{Performance Prediction Recipes for Optical Links}
\author{Erik Agrell, \emph{Fellow, IEEE,} Marco Secondini, \emph{Senior Member, IEEE,} Alex Alvarado, \emph{Senior Member, IEEE,} and Tsuyoshi Yoshida, \emph{Member, IEEE}
\thanks{This research was supported in part by
the Swedish Research Council (VR) (grant no.~2017-03702),
the Knut and Alice Wallenberg Foundation (grant no.~2018.0090), 
the European Research Council (ERC) under the European Union's Horizon 2020 research and innovation programme (grant agreement no.~757791),
and
National Institute of Information and Communications Technology (grant no.~20401).}
\thanks{E.~Agrell is with the Department of Electrical Engineering, Chalmers University of Technology, 41296 Gothenburg, Sweden.}
\thanks{M.~Secondini is with the Institute of Communication, Information and Perception Technologies,
Scuola Superiore Sant'Anna, Pisa, Italy, and with the National Laboratory of Photonic Networks, CNIT, Pisa, Italy.}
\thanks{A.~Alvarado is with the Information and Communication Theory Lab, Signal Processing Systems Group, Department of Electrical Engineering, Eindhoven University of Technology, 5600 MB, Eindhoven, The Netherlands.}
\thanks{T.~Yoshida is with Information Technology R\&D Center, Mitsubishi Electric Corporation, 5-1-1 Ofuna, Kamakura, 247-8501, Japan, and with Graduate School of Osaka University, 2-1 Yamadaoka, Suita, 565-0871, Japan.}
}

\IEEEoverridecommandlockouts %
\IEEEspecialpapernotice{(Invited Tutorial)}

\begin{document}

\maketitle

\begin{abstract}
Although forward error-correction (FEC) coding is an essential part of modern fiber-optic communication systems, it is impractical to implement and evaluate FEC in transmission experiments and simulations. Therefore, it is desirable to accurately predict the end-to-end link performance including FEC from transmission data recorded without FEC. In this tutorial, we provide ready-to-implement ``recipes'' for such prediction techniques, which apply to arbitrary channels and require no knowledge of information or coding theory. The appropriate choice of recipe depends on properties of the FEC encoder and decoder. The covered metrics include bit error rate, symbol error rate, achievable information rate, and asymptotic information, in all cases computed using a mismatched receiver. Supplementary software implementations are available.
\end{abstract}

\begin{IEEEkeywords}
Achievable information rate, asymmetric information, bit error rate, mismatched decoding, optical communication, performance metrics, probabilistic shaping.
\end{IEEEkeywords}

\section{Introduction} \label{sec:intro}

Forward error-correction (FEC) is ubiquitous in practically all deployed fiber-optical communication systems. By adding a controlled amount of overhead on top of the data payload, the receiver can correctly decode the data even in the presence of channel noise and other impairments, provided that the coding scheme and overhead are suitably chosen. A common target is a bit error rate (BER) of $10^{-15}$ after FEC (post-FEC BER), which for a $100$~Gbit/s link means about $9$ bit errors per day. It may be very time-consuming and costly to reliably estimate even a single BER value in this ultralow regime. Doing it multiple times in order to test different configurations or fine-tune  parameters is out of the question. Thus, system experiments and simulations are usually carried out without FEC. The topic of this brief tutorial is how to predict the performance of systems \emph{with FEC} from performance metrics for systems \emph{without FEC.}

The history of FEC and coded modulation for optical systems begins with \emph{hard-decision} (HD) binary block codes and binary modulation. For such codes, including the standardized $(255, 239)$ Reed--Solomon code \cite{itu00}, theoretical approximations %
and tables accurately describe the relations between pre-FEC BER and post-FEC BER \cite{itu00,itu04}. The pre-FEC BER also offers good performance prediction in systems with \emph{bit-interleaved coded modulation} \cite{szczecinski15}%
, still under HD bit-wise decoding \cite{schmalen17}. In systems with \emph{soft-decision} (SD) FEC decoding \cite{leven11,alvarado15} and \emph{probabilistic shaping} (PS) \cite{cho19, yoshida20benchmarking}, however, information-theoretic metrics are needed for accurate performance prediction. 

A good pre-FEC performance metric is one that has a deterministic, monotonic relation with the post-FEC BER for a given FEC code, a relation that should ideally remain the same for a wide range of channels, modulation formats, and signal processing algorithms. With such a metric, alternative link configurations can be compared without implementing any FEC, and the conclusions would be valid also with FEC. Furthermore, the pre-FEC metric can be compared with a precomputed threshold to predict whether the post-FEC BER would satisfy the requirement or not. Common pre-FEC metrics are here computed using \emph{recipes,} (available at \href{http://codes.se/software}{\emph{codes.se/software}}),
which apply to arbitrary channel models or experimental data. Importantly, they are not restricted to the additive white Gaussian noise (AWGN) channel as in some previous works. %

\section{Data Generation and Processing} \label{sec:data}

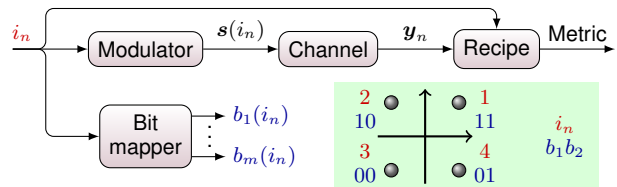
\begin{figure}
\begin{center}
\tikzstyle{box} = [rectangle, fill=white, rounded corners, draw, text centered, inner sep=4pt, top color=white,
bottom color=purple!50!black!20]
\tikzstyle{arrow} = [-latex, rounded corners=3pt]
\sffamily\small
\colorlet{figred}{red!80!black}
\colorlet{figblue}{blue!60!black}
\colorlet{figgray}{black!40!white}
\colorlet{figgreen}{green!15!white}

\begin{tikzpicture}[scale=0.9, every node/.style={scale=0.9}]
\node (leftend) {};
\node [box, right=10mm of leftend] (mod) {Modulator};
\node [box, right=10mm of mod] (cha) {Channel};
\node [box, right=10mm of cha] (rec) {Recipe};
\node [box, below=5mm of mod, align=center] (map) {Bit\\mapper};
\node [right=10mm of rec] (rightend) {};
\draw [arrow] (leftend) -- node[above, very near start, color=figred]{$i_n$} (mod);
\draw [arrow] (mod) -- node[above]{$\bs(i_n)$} (cha);
\draw [arrow] (cha) -- node[above]{$\by_n$} (rec);
\draw [arrow] (rec) -- node[above]{Metric} (rightend);
\draw[arrow] (leftend) -- ++(6mm,0) -- ++(0,6mm) -| (rec.north); 

\draw[arrow] (leftend) -- ++(6mm,0) -- ++(0,-6mm) |- (map.west); 
\draw [arrow] (map.east)++(0,3mm) -- +(5mm,0) node[right, color=figblue]{$b_1(i_n)$} (rightend);
\draw [arrow] (map.east)++(0,-3mm) -- +(5mm,0) node[right, color=figblue]{$b_m(i_n)$} (rightend);
\node [right=1mm of map, yshift=1mm, align=center] {$\vdots$};

\newcommand{\dx}{5mm}
\node [right=30mm of map.east] (origin) {};
\path[fill=figgreen] (map.east)+(21mm,-8mm) rectangle +(60mm,8mm);
\draw [->, thick] (origin)+(-7mm,0) -- +(7mm,0);
\draw [->, thick] (origin)+(0,-7mm) -- +(0,7mm);
\draw[ball color=figgray] (origin)+(-\dx,\dx) node[left=1mm, align=center, yshift=-1mm]{{\color{figred} $2$}\\{\color{figblue} $10$}} circle(1mm);
\draw[ball color=figgray] (origin)+(-\dx,-\dx) node[left=1mm, align=center, yshift=1mm]{{\color{figred} $3$}\\{\color{figblue} $00$}} circle(1mm);
\draw[ball color=figgray] (origin)+(\dx,-\dx) node[right=1mm, align=center, yshift=1mm]{{\color{figred} $4$}\\{\color{figblue} $01$}} circle(1mm);
\draw[ball color=figgray] (origin)+(\dx,\dx) node[right=1mm, align=center, yshift=-1mm]{{\color{figred} $1$}\\{\color{figblue} $11$}} circle(1mm);
\node [right=14mm of origin, align=center] {{\color{figred} $i_n$}\\{\color{figblue} $b_1 b_2$}};
\end{tikzpicture}

\caption{\label{fig:pairs}Data generation, performance metric computation, and a simple bit mapping example (4-QAM).}
\end{center}
\end{figure}

The performance metrics are computed from a set of input and output
data, which are generated according to the scheme in Fig.~\ref{fig:pairs}.
The constellation consists of $M$ symbols, represented as real-valued vectors, whose dimension $D=2$ in the common case of quadrature amplitude modulation (QAM). Otherwise $D=1$ for pulse amplitude modulation or $D>2$ if multiple modes (in polarization, wavelength, time, and/or space) are jointly modulated and demodulated.
Each symbol is
associated with an index $1\le i\le M$
and $m=\log_{2}M$ bits $b_{1}(i),\ldots,b_{m}(i)$. An example with
$D=2$, $M=4$, and Gray labeling is illustrated in Fig.~\ref{fig:pairs}.

Throughout this paper, we are interested in FEC performance prediction for a sequence of $N$ symbols.
The data generation process starts with drawing $N$
independent and identically distributed indices $i_{1},\ldots,i_{N}$ from a given probability distribution $p_j$, which is traditionally uniform ($p_j=1/M$, Sec.~\ref{sec:hard}--\ref{sec:soft}) and nonuniform in systems with PS (Sec.~\ref{sec:ps}). 
The indices are then mapped to the corresponding symbols $\bs(i_{1}),\ldots,\bs(i_{N})$ in the modulator
and fed to the channel, which produces the output symbols
$\by_{1},\ldots,\by_{N}$. In general, the channel represents
either a physical channel (as in a lab experiment) or its numerical
emulation (as in a computer simulation) and must include all relevant
analog and digital operations (e.g., filtering, D/A and A/D conversion,
and synchronization, but no FEC encoding/decoding) performed at the transmitter and at the receiver.

As discussed in the next sections, the relevant performance metric
for the considered system depends on the decoding strategy (SD,
HD, bit-wise, symbol-wise, \ldots) employed at the receiver. In an
optimal receiver, the decoding strategy should also account for the
exact input--output relation of the channel\new{, which is nonlinear and
bursty \cite{agrell18ipc}.} Since \new{this exact relation} is
generally unknown, a \emph{mismatched} receiver---designed according
to a simplified (auxiliary) channel model---is usually employed.
Typically, the receiver is designed by assuming that the output samples
are simply corrupted by AWGN with a variance per dimension equal to
\begin{equation}
\sigma^{2}=\frac{1}{DN}\sum_{n=1}^{N}\left\Vert \by_{n}-\bs(i_{n})\right\Vert ^{2}.
\label{eq:sigma}
\end{equation}
We emphasize that this is not the actual behavior of the channel and
is just the implicit assumption made, for instance, when using a
Euclidean distance metric in the decoder. In this case, it is fundamental
that the channel includes all the processing blocks that are needed
to minimize the mismatch between the true channel and the adopted
model, such as trivial scaling, rotation, and delay as well as
\new{symbol interleaving to disperse error bursts and}
more refined algorithms for clock recovery, carrier recovery, and channel equalization.
\new{The recipes in the following sections provide accurate performance predictions for matched as well as mismatched receivers \cite{schmalen17,liga17,secondini19}.}

\section{Recipes with HD FEC} \label{sec:hard}

Hard decisions means that the receiver takes a tentative decision $\hi_n$ on the transmitted data $i_n$ corresponding to each received vector $\by_n$. This data $\hi_n$ is then fed into the HD FEC decoder for possible error correction. The most common HD rule is the minimum Euclidean distance
\begin{align}
\hi_n &= \argmin_{j\in\{1,2,\ldots,M\}} \|\by_n - \bs(j) \|^2 \label{eq:ihat},
\end{align}
which implicitly assumes a mismatched receiver optimized for the memoryless AWGN channel.

In HD systems, two standard pre-FEC metrics are the \emph{symbol error rate} (SER) $P_\mathrm{s}$ and BER $P_\mathrm{b}$, which simply count errors between the transmitted and received data:
\begin{align}
P_\mathrm{s} &= \frac{1}{N}\sum_{n=1}^N \delta\!\left( i_n \ne \hi_n \right) \label{eq:ser},\\
P_\mathrm{b} &= \frac{1}{mN}\sum_{n=1}^N\sum_{k=1}^m\delta\!\left( b_k(i_n) \ne b_k(\hi_n) \right) \label{eq:ber}.
\end{align}
The function $\delta(\cdot)$ is $1$ if the argument is true and $0$ otherwise.

The SER is, albeit conceptually very simple, not very useful in practice. It is relevant for nonbinary coding using a coding alphabet of the same size as the constellation and the Hamming distance as the FEC decoding metric. AIRs for such systems are discussed in \cite{liga17,sheikh17}. However, nonbinary coded modulation is not very common in practice, since its performance gains over BICM are typically modest compared with the significantly increased complexity \cite[Sec.~7.6.4]{graell20}.

The BER is a relevant performance metric for BICM with binary FEC and HD decoding, which is a very common setup in deployed systems. If the BER \eqref{eq:ber} is below a certain threshold, which varies depending on the selected code, then a satisfactory post-FEC performance is guaranteed, provided that sufficiently deep bit-interleaving is applied. Such BER thresholds have been tabulated for many binary codes \cite{agrell18ipc,graell20} of various code rates $\Rc$. BER results and BER thresholds are often equivalently presented in terms of the \emph{Q factor} $Q_{\mathrm{hard}} = \sqrt{2}\mathrm{erfc}^{-1} (2 P_\mathrm{b})$, which gives the theoretical signal-to-noise ratio (SNR) of a binary-input AWGN channel to achieve the same BER $P_\mathrm{b}$.

The AIR using a selected code for which the BER condition is met equals $m \Rc$ bit/symbol. If on the other hand ideal FEC coding is assumed, then the AIR with HD BICM is 
\begin{align}\label{eq:airb_HD}
\mathrm{AIR}_\mathrm{b}^\mathrm{HD}= m(1-H_2(P_\mathrm{b})) \ge m \Rc,
\end{align}
where $H_2$ is the binary entropy function \cite[Eq.~(25)]{liga17}.

\section{Recipes with SD FEC} \label{sec:soft}

In this section, we discuss soft-decision FEC, where the receiver does not take a decision on the transmitted index (as in \eqref{eq:ihat}), but instead, it uses \emph{soft information} (reliability information) available in the received sequence $\by_n$. Here we consider two AIRs, which are valid for systems with symbol-wise and bit-wise receivers, resp., namely \new{\cite[Eq.~(52)--(53)]{alvarado18}}
\begin{align}
\nonumber
\mathrm{AIR}_\mathrm{s} &= m-\frac{1}{N}\sum_{n=1}^N \biggl\{\log_2 {\sum_{j=1}^{M} q(\by_n,\bs(j))}\\
&\qquad\qquad\qquad\qquad\quad- \log_2{q(\by_n,\bs(i_n))}\biggl\} , \label{eq:airs}\\
\nonumber
\mathrm{AIR}_\mathrm{b} &= m-\frac{1}{N}\sum_{n=1}^N\sum_{k=1}^m \biggl\{\log_2 {\sum_{j=1}^{M} q(\by_n,\bs(j))} \\ 
&- \log_2 {\sum_{j=1}^{M} \delta(b_k(j)=b_k(i_n)) q(\by_n,\bs(j))}\biggr\} \label{eq:airb},
\end{align}
where
\begin{align}
q(\by,\bs) &= \exp\!\left(-\|\by-\bs\|^2/(2\sigma^2)\right) \label{eq:q}
\end{align}
and $\sigma^2$ is given by \eqref{eq:sigma}.
 
When compared to the SER and BER recipes in \eqref{eq:ser} and \eqref{eq:ber}, the AIRs in \eqref{eq:airs} and \eqref{eq:airb} only require one additional summation, and thus, their computation is still very simple. We emphasize again that the choice of $q(\by,\bs)$ in \eqref{eq:q} is arbitrary and might not correspond to the actual behavior of the channel under consideration. More details are discussed in Sec.~\ref{sec:discussion}.

Like the SER, $\mathrm{AIR}_\mathrm{s}$ is less popular because it applies to systems based on nonbinary codes \cite{schmalen17}, and in practice binary codes are preferred. $\mathrm{AIR}_\mathrm{b}$ is a relevant performance metric for BICM with binary FEC and SD decoding. Such systems are popular in practice because of the ease of implementation of binary codes and the improved performance over HD-FEC. 

$\mathrm{AIR}_\mathrm{b}$ is an AIR with BICM and ideal binary SD-FEC. With a practical FEC, the actual rate is $m R_c\le \mathrm{AIR}_\mathrm{b}$, while $\mathrm{AIR}_\mathrm{b}$ can be used to predict the post-FEC BER.
When $\mathrm{AIR}_\mathrm{b}/m$ is below a certain threshold, it can be claimed that a certain post-FEC BER will be obtained at a given coding rate $R_{\mathrm{c}}$. The exact value of the threshold depends on the specific code under consideration. Thresholds for $\mathrm{AIR}_\mathrm{b}/m$ for a concatenated FEC scheme with various inner SD-FEC codes and an outer HD-FEC code (not simulated) were first published in \cite[Table~III]{alvarado15}. Recently, thresholds for SD-FEC for an approximated post-FEC BER of $10^{-15}$ have been reported in \cite[Table~7.5]{graell20}. An $\mathrm{AIR}_{\mathrm{b}}$ can be converted to a Q factor as $Q_{\mathrm{soft}} = \sqrt{J^{-1} ( \mathrm{AIR_{\mathrm{b}}}/m )/2}$, where $J$ is defined in \cite[App.]{tenbrink04}. This is the SD analogy of $Q_\mathrm{hard}$ in Sec.~\ref{sec:hard}.

The metrics discussed in this section are applicable to soft-decision FEC where the bits and symbols are equally likely. The soft information passed to the decoder is typically represented using \emph{L-values} (or log-likelihood ratios). These L-values can also consider systems where bits or symbols are transmitted with different probabilities, which is the main focus of the next section.

\section{Recipes with Probabilistic Shaping} \label{sec:ps}

The most popular PS scheme in recent years is probabilistic amplitude shaping,
where the 
shaping operation, often realized using \emph{distribution matching} (DM), is placed outside the FEC coding operation. The DM encodes on average $R_\mathrm{p}$ bits onto each shaped symbol, where $R_\mathrm{p}$ is less than the symbol entropy $H_\mathrm{s}=-\sum_{j=1}^M p_j \log_2 p_j$ \cite{bocherer19}. 
This system uses binary FEC and memoryless bit-wise decoding. When employing SD FEC, bit-wise %
L-values
\begin{align}
L_{n,k} &= %
\ln \frac{\sum_{j=1}^{M} \delta(b_k(j)=0) p_j q(\by_n,\bs(j))}{\sum_{j=1}^{M} \delta(b_k(j)=1)  p_j q(\by_n,\bs(j))} , \label{eq:Lval} %
\end{align}
with $q$ given in \eqref{eq:q},
are fed into the FEC decoder for every received vector $\by_n$.
Several performance metrics for the post-FEC BER of systems with PS and SD FEC have been proposed \cite{yoshida17,bocherer19,cho19,yoshida20benchmarking}. In this short tutorial, we give a recipe for one of them, the \emph{asymmetric information} (ASI). To estimate the ASI,
we first estimate the distribution of the \emph{asymmetric L-values} $L^\mathrm{a}_{n,k} = (-1)^{b_k(i_n)} L_{n,k}$ via a histogram. For a given number of bins $B$ with separation $\Delta$, we define a quantization function $T(\ell) = \argmin_j | \ell-\ell_j |$ where $\ell_j = (2j-1-B)\Delta$ for $j=1,\ldots,B$. Using the probability estimates 
\begin{align} \label{eq:histogram}
\Lambda_j = \frac{1}{mN} \Big| \Big\{ T(L^\mathrm{a}_{n,k}) = \ell_j : 1\le n \le N, 1 \le k \le m \Big\} \Big|
\end{align}
for $j = 1,\ldots, B$, the ASI estimate is
\begin{align} \label{eq:asi}
\mathrm{ASI} &= \sum_{\substack{j=1, \Lambda_j \neq 0}}^B \Lambda_j \log_2 \left( \frac{2\Lambda_j}{\Lambda_j+\Lambda_{B+1-j}} \right) .
\end{align}
For sufficiently high $N$ and suitable choices of $B$ and $\Delta$, the estimate \eqref{eq:histogram}--\eqref{eq:asi} approaches the true ASI \cite[Eq.~(35)]{yoshida20benchmarking}. We suggest $B=32$ and $\Delta=1$ as default parameters.\footnote{Larger or smaller $\Delta$ may be beneficial at very high or low SNRs, resp.}

The same thresholds as for $\mathrm{AIR}_\mathrm{b}/m$ in Sec.~\ref{sec:soft} are valid for $\mathrm{ASI}$, because the $\mathrm{ASI}$ is an %
achievable binary FEC code rate for uniform and PS signals and can be used to predict the post-FEC BER \cite{yoshida17}.
Similarly to the case in Sec.~\ref{sec:soft}, an ASI can be converted into a Q factor as $Q_{\mathrm{soft}} = \sqrt{J^{-1} ( \mathrm{ASI} )/2}$. %
An AIR with BICM and ideal binary SD-FEC and DM is 
\begin{align} \label{eq:air_ps}
\mathrm{AIR}_{\mathrm{b}}^{\mathrm{PS}} &= H_\mathrm{s} - (1 - \mathrm{ASI}) m .
\end{align}
With practical (nonideal) FEC and %
DM, the net data rate is $R_\mathrm{p}-(1-R_\mathrm{c})m\le \mathrm{AIR}_{\mathrm{b}}^\mathrm{PS}$. %

A suitable metric for HD PS systems with bit-wise decoding is the pre-FEC BER \cite[Eq.~(30)]{yoshida20benchmarking}, which can be numerically estimated as
\begin{align}
P_\mathrm{b}^{\mathrm{PS}} &= \frac{1}{mN} \sum_{n=1}^{N} \sum_{k=1}^{m} \delta( L^\mathrm{a}_{n,k} \le 0 ) . \label{eq:preber_ps} %
\end{align}
It determines the maximum AIR of such systems \cite[Sec.~II]{sheikh18}. In the special case of $p_j = 1/M$, $P_\mathrm{b}^{\mathrm{PS}}$ is similar to \eqref{eq:ber} for many channels, especially at high SNR, but not identical, since \eqref{eq:preber_ps} applies bit-wise decisions \cite{ivanov13} and \eqref{eq:ber} symbol-wise.

\section{Examples} \label{sec:examples}

Here we consider a memoryless channel with both phase noise and AWGN, defined by $\by_n = \bR(\theta_n)\bs(i_n) + \bZ_n$, which has been considered in many previous works \cite{Koike-AkinoCLEO15}. Here $D=2$, the matrix $\bR(\theta_n)$ denotes a rotation by a random angle $\theta_n$, which is zero-mean Gaussian with variance $\sigma^2_{\theta}$, and $\bZ_n$ is independent AWGN with variance $\sigma_z^2$ in each dimension.

\begin{figure}
\centering
\colorlet{mygreen}{green!50!black}
\colorlet{myblue}{blue!70!white}
\begin{tikzpicture}[tight background]
\begin{semilogyaxis}[
	width=1.0\columnwidth,
	height=0.65\columnwidth,
	every axis/.append style={font=\footnotesize},
	ylabel={Error Probability}, 
	xlabel= {AWGN SNR~[dB]},
	xmin=5,
	xmax=30,
	ymin=1e-3,
	ymax=1,
	grid=both,
	ylabel style={yshift=-0.25cm},
	xlabel style={yshift=+0.15cm},
	xminorgrids=true,
	legend style={legend style={at={(.00,.02)},anchor=south west,draw=none},font=\ssmall,legend cell align=left,row sep=-0.15em,name=legend,inner xsep=2pt,inner ysep=0pt}]
	\addplot+[myblue,solid,mark=*,mark size=1.5] file {P_s_64QAM_sigmap2_0.01.txt};
		\addlegendentry{$P_{\mathrm{s}}$ \eqref{eq:ser}, 64-QAM};
	\addplot+[myblue,solid,mark=*,mark size=1.5,mark options={fill=white,solid}] file {P_b_64QAM_sigmap2_0.01.txt};
		\addlegendentry{$P_{\mathrm{b}}$ \eqref{eq:ber}, 64-QAM};
	\addplot+[mygreen,solid,mark=*,mark size=1.5,mark options={fill=white,solid}] file {Pbps_PS256QAM_sigmap2_0.01.txt};
		\addlegendentry{$P_{\mathrm{b}}^{\mathrm{PS}}$ \eqref{eq:preber_ps}, PS, $q$ in \eqref{eq:q}};
	\addplot+[mygreen,densely dashed,mark=*,mark size=1.5,mark options={fill=white,solid}] file {Pbps_BLT_PS256QAM_sigmap2_0.01.txt};
		\addlegendentry{$P_{\mathrm{b}}^{\mathrm{PS}}$ \eqref{eq:preber_ps}, PS, $q$ in \cite[Eq.~(3)]{Koike-AkinoCLEO15}};
\end{semilogyaxis}
\end{tikzpicture}
\begin{tikzpicture}[tight background]
\begin{axis}[
	width=1.05\columnwidth,
	height=0.65\columnwidth,
	every axis/.append style={font=\footnotesize},
	ylabel={Achievable Rate [bit/symbol]},
	xlabel= {AWGN SNR~[dB]},
	xmin=5,
	xmax=30,
	ymin=0.75,
	ymax=6.3,
	grid=both,
	minor y tick num=1,
	ylabel style={yshift=-0.15cm},
	xlabel style={yshift=+0.15cm},
	legend style={legend style={at={(1.00,.02)},anchor=south east,draw=none},font=\ssmall,legend cell align=left,row sep=-0.15em,name=legend,inner xsep=2pt,inner ysep=0pt}]
	\addplot+[myblue,mark=*,mark size=1.5,mark repeat=2,mark phase=1,mark options={fill=white,solid}] file {AIR_b_HD_64QAM_sigmap2_0.01.txt};
		\addlegendentry{$\mathrm{AIR}_\mathrm{b}^\mathrm{HD}$ \eqref{eq:airb_HD}, 64-QAM},
	\addplot+[myblue,mark=square*,mark size=1.0,mark repeat=2,mark phase=2,mark options={fill,solid}] file {AIR_s_64QAM_sigmap2_0.01.txt};
		\addlegendentry{$\mathrm{AIR}_\mathrm{s}$ \eqref{eq:airs}, 64-QAM, $q$ in \eqref{eq:q}},
 	\addplot+[myblue,mark=square*,mark size=1.0,mark repeat=2,mark phase=1,mark options={fill=white,solid}] file {AIR_b_64QAM_sigmap2_0.01.txt};	
		\addlegendentry{$\mathrm{AIR}_\mathrm{b}$ \eqref{eq:airb}, 64-QAM, $q$ in \eqref{eq:q}},
 	\addplot+[mygreen,mark=square*,mark size=1.0,mark repeat=2,mark phase=2,mark options={fill=white,solid}] file {AIRps_SD_PS256QAM_sigmap2_0.01.txt};
		\addlegendentry{$\mathrm{AIR}_\mathrm{b}^\mathrm{PS}$ \eqref{eq:air_ps}, PS, $q$ in \eqref{eq:q}},
 	\addplot+[myblue,densely dashed,mark=square*,mark size=1.0,mark repeat=2,mark phase=1,mark options={fill,solid}] file {AIR_s_BLT_64QAM_sigmap2_0.01.txt};
		\addlegendentry{$\mathrm{AIR}_\mathrm{s}$ \eqref{eq:airs}, 64-QAM, $q$ in \cite[Eq.~(3)]{Koike-AkinoCLEO15}},
 	\addplot+[myblue,densely dashed,mark=square*,mark size=1.0,mark repeat=2,mark phase=2,mark options={fill=white,solid}] file {AIR_b_BLT_64QAM_sigmap2_0.01.txt};	
		\addlegendentry{$\mathrm{AIR}_\mathrm{b}$ \eqref{eq:airs}, 64-QAM, $q$ in \cite[Eq.~(3)]{Koike-AkinoCLEO15}},
	\addplot+[mygreen,densely dashed,mark=square*,mark size=1.0,mark repeat=2,mark phase=1,mark options={fill=white,solid}] file {AIRps_SD_BLT_PS256QAM_sigmap2_0.01.txt};
		\addlegendentry{$\mathrm{AIR}_\mathrm{b}^\mathrm{PS}$ \eqref{eq:air_ps}, PS, $q$ in \cite[Eq.~(3)]{Koike-AkinoCLEO15}}, 	
\end{axis}
\end{tikzpicture}
\caption{Error probabilities (top) and AIRs (bottom).}
\label{fig:example}
\end{figure}
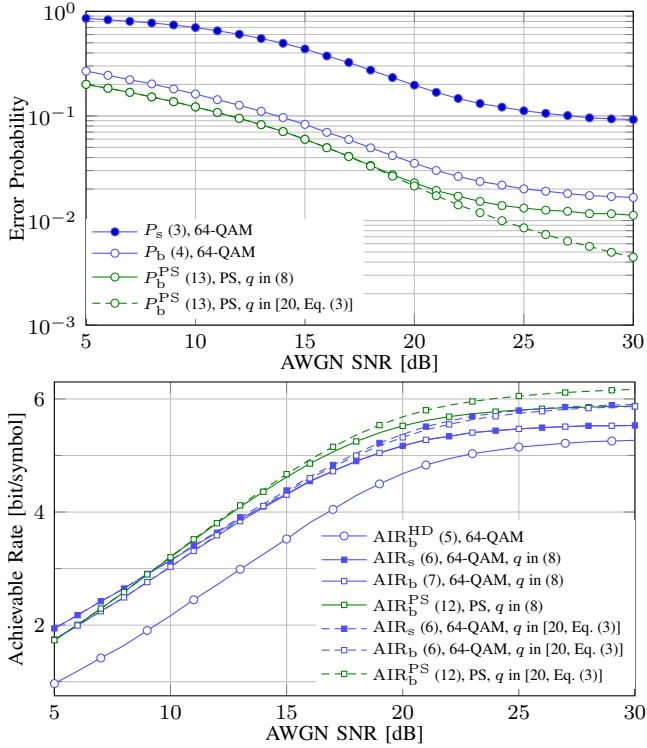

We use $N=10^6$ samples \new{to obtain reasonably accurate estimates} and $\sigma_\theta^2=0.01$ \new{to make the impact of phase noise clearly visible in the considered SNR range.}
We study uniform $64$-QAM (%
blue, $H_\mathrm{s}=6$~bit/symbol) and PS $256$-QAM (%
green, $H_\mathrm{s}=6.3$~bit/symbol), both with Gray bit mapping. Fig.~\ref{fig:example} (top) shows the SER and BER estimates %
as functions of the AWGN SNR, defined as the signal variance (energy) per dimension divided by $\sigma_z^2$. These results show that even at high SNRs, %
$P_{\mathrm{b}}$ and $P_{\mathrm{s}}$ reach error floors. This is due to the fact we use \eqref{eq:ihat}, which is a suboptimal decision rule for this channel.
$P_{\mathrm{b}}^{\mathrm{PS}}$ with the mismatched $q(\by,\bs)$ in \eqref{eq:q} behaves similarly.

Fig.~\ref{fig:example} (bottom) %
shows that the AIR $\mathrm{AIR}_\mathrm{b}^\mathrm{HD}$ in \eqref{eq:airb_HD} is considerably lower than all other AIRs, which is due to the HD-FEC assumption. Fig.~\ref{fig:example} (bottom) shows that $\mathrm{AIR}_\mathrm{s}\approx \mathrm{AIR}_\mathrm{b}$ for AIRs above 4~bit/symbol, due to the use of a Gray labeling. Similarly to Fig.~\ref{fig:example} (top),  
$\mathrm{AIR}_\mathrm{s}$, $\mathrm{AIR}_\mathrm{b}$, and $\mathrm{AIR}_\mathrm{b}^\mathrm{PS}$
saturate at high SNRs, which in this case is due to the use of the mismatched $q(\by,\bs)$ in \eqref{eq:q}. To improve these AIRs, we use \cite[Eq.~(3)]{Koike-AkinoCLEO15},%
\footnote{Specifically, we set $\log q(\by,\bs)$ equal to the right-hand side of \cite[Eq.~(3)]{Koike-AkinoCLEO15}, where $R=(3y_{c}-s_{c})/2$, $S=(y_{c}+s_{c})/2$, and $y_{c}$ and $s_{c}$ are the complex symbols whose real and imaginary parts are the two components of $\by$ and $\bs$.} 
which is a good approximation for the channel under consideration. Here both $\sigma_\theta^2$ and $\sigma_z^2$ are assumed to be perfectly estimated. This $q(\by,\bs)$ is better matched to the channel, which results in lower $P_{\mathrm{b}}^{\mathrm{PS}}$ and higher AIRs in Fig.~\ref{fig:example} (dashed). %

\section{Extensions and Conclusions} \label{sec:discussion}

Many variants of the presented metrics exist. The recipes \eqref{eq:ser}--\eqref{eq:airb}, \eqref{eq:asi}, and \eqref{eq:preber_ps} give valid performance metrics for arbitrary $\hi$ and arbitrary positive $q$, not only $\eqref{eq:ihat}$ and $\eqref{eq:q}$, as exemplified in Sec.~\ref{sec:examples}. These metrics are achievable by receivers that use the same $\hi$ and $q$. Improved performance is often attained by optimization over various parameters. For example, $\sigma^2$ in \eqref{eq:q} can be optimized, which is analogous to optimization over the exponent $s$ in \cite[Eq.~(4.34)]{szczecinski15}, \cite[Example~2]{schmalen17}. Other options include, e.g., to let $\sigma^2 = \sigma^2_i$ depend on the symbol index $i$, to define $q$ according to a correlated, possibly multidimensional, Gaussian distribution \cite{eriksson16}, and to capture channel memory via hidden Markov models \cite{secondini19}.

To conclude, the presented practical recipes are related to well-known information-theoretic quantities. If \eqref{eq:airs} is optimized over all positive functions $q$, which for general channels is possible only in theory, then the resulting AIR approaches the memoryless \emph{mutual information} as $N\rightarrow\infty$. If \eqref{eq:airb} is similarly optimized over $q$, then the \emph{generalized mutual information} for BICM \cite[Eq.~(4.53)]{szczecinski15} is obtained. If the mutual information is further optimized over all possible constellations $\bs(i)$ and/or distributions $p_j$ while $M\rightarrow\infty$, which corresponds to optimizing over all possible input distributions, then the \emph{channel capacity} is obtained. Therefore, the AIRs in this paper are often presented as \emph{capacity lower bounds.}

\end{document}